\documentclass[prd, superscriptaddress, tightenlines, longbibliography, nofootinbib, eqsecnum, amsfonts, amsmath, floatfix, notitlepage]{revtex4-1}
\pdfoutput=1

\usepackage[utf8]{inputenc}
\usepackage{mathrsfs}
\usepackage{euscript}
\usepackage{epsfig}
\usepackage{graphics}
\usepackage{graphicx}
\usepackage{amsmath}
\usepackage{amssymb}
\usepackage{bm}
\usepackage[usenames,dvipsnames,svgnames,table]{xcolor}
\usepackage{xspace}
\usepackage{wasysym}
\usepackage{times}
\usepackage{appendix}
\usepackage{lipsum}
\usepackage[nolist,nohyperlinks]{acronym}
\usepackage{float}
\usepackage{simplewick}
\usepackage{natbib, ifthen}
\usepackage{hyperref}

\defcitealias{Pratten:2016dsm}{PL16}
\newcommand{\paperone}{\citetalias{Pratten:2016dsm}}

\newcommand{\la}{\langle}
\newcommand{\ra}{\rangle}

\newcommand{\mksym}[1]{\ifmmode {\rm #1}\else #1\fi}

\providecommand{\lea}{\la}
\providecommand{\gea}{\ga}

\providecommand{\alt}{\lea}
\providecommand{\agt}{\gea}
\providecommand{\text}[1]{\rm{#1}}

\newcommand{\grad}{\nabla}

\providecommand{\HALOFIT}{{\tt halofit}}

\newcommand{\begm}{\begin{pmatrix}}
\newcommand{\enm}{\end{pmatrix}}

\newcommand\ba{\begin{eqnarray}}
\newcommand\ea{\end{eqnarray}}
\newcommand\bea{\begin{eqnarray}}
\newcommand\eea{\end{eqnarray}}

\newcommand\be{\begin{equation}}
\newcommand\ee{\end{equation}}

\newcommand{\valpha}{{\boldsymbol{\alpha}}}
\newcommand{\vgrad}{{\boldsymbol{\nabla}}}

\newcommand{\vell}{{\boldsymbol{\ell}}}

\newcommand{\ud}{{\rm d}}

\newcommand{\mA}{\bm{A}}

\newcommand{\boldvec}[1]{{\mbox{\boldmath{$#1$}}}}

\newcommand{\vL}{\boldvec{L}}

\providecommand{\vr}{\boldvec{r}}

\newcommand{\vx}{\boldvec{x}}

\newcommand{\clo}{\mathcal{O}}

\newcommand{\vrhat}{\hat{\vr}}

\newcommand{\trig}{\text{trig}}

\renewcommand{\vr}{\boldvec{r}}
\newcommand{\Cgtwo}{C_{\rm{gl},2}}
\newcommand{\vellhat}{\hat{\vell}}
\newcommand{\lgl}{\vellhat^T\gamma\vellhat}

\definecolor{ZurichBlue}{rgb}{.255,.41,.884} 		
\definecolor{ZurichRed}{rgb}{0.9, 0.1, 0} 			
\definecolor{ZurichGreen}{rgb}{.196,.504,.396} 		
\definecolor{ZurichYellow}{rgb}{1,.648,0} 			
\definecolor{dodgerblue}{rgb}{0.12, 0.56, 1.0}
\definecolor{azure}{rgb}{0.0, 0.5, 1.0}
\definecolor{awesome}{rgb}{1.0, 0.13, 0.32}
\definecolor{alizarincrimson}{rgb}{0.82, 0.1, 0.26}
\definecolor{mediumpurple}{rgb}{0.58, 0.44, 0.86}
\definecolor{lasallegreen}{rgb}{0.03, 0.47, 0.19}

\DeclareMathAlphabet{\pazocal}{OMS}{zplm}{m}{n}

\begin{document}

\newcommand{\ie}{i.e.}
\newcommand{\etal}{\textit{et al.}}
\newcommand{\rmi}{{\rm i}}

\newcommand{\ellp}{\ell^{\prime}}
\newcommand{\bfell}{{\pmb{\ell}}}
\newcommand{\bfellp}{{\pmb{\ell}^{\prime}}}
\newcommand{\bfellpp}{{\pmb{\ell}^{\prime \prime}}}

\newcommand{\alphav}{{\boldsymbol{\alpha}}}
\newcommand{\bfA}{{\bf{A}}}
\newcommand{\bfe}{{\bf{e}}}
\newcommand{\bfr}{{\bf{r}}}
\newcommand{\bfn}{{\bf{n}}}
\newcommand{\bfnp}{{\bf{n}}^{\prime}}
\newcommand{\bfx}{{\bf{x}}}
\newcommand{\bfk}{{\bf{k}}}
\newcommand{\bfd}{{\bf{d}}}
\newcommand{\bfC}{{\bf{C}}}
\newcommand{\bfL}{{\bf{L}}}
\newcommand{\bfLp}{{\bf{L}}^{\prime}}
\newcommand{\bfLpp}{{\bf{L}}^{\prime \prime}}
\newcommand{\bfLppp}{{\bf{L}}^{\prime \prime \prime}}

\newcommand{\bfmu}{{\bm\mu}}
\newcommand{\bftgr}{{\bm\theta}^{\textrm{GR}}}
\newcommand{\bftmg}{{\bm\theta}^{\textrm{MG}}}
\newcommand{\bft}{{\bm\theta}}
\newcommand{\bfa}{{\bm\alpha}}
\newcommand{\bftp}{{\bm\theta}^{\prime}}

\newcommand{\calA}{{\mathcal{A}}}
\newcommand{\calC}{{\mathcal{C}}}
\newcommand{\calD}{{\mathcal{D}}}
\newcommand{\calR}{{\mathcal{R}}}
\newcommand{\calO}{{\mathcal{O}}}
\newcommand{\calS}{{\mathcal{S}}}

\newcommand{\dlw}{{(2 \pi)^2}}

\newcommand{\lef}{{g (\chi , \chi^{\prime})}}
\newcommand{\chip}{{\chi^{\prime}}}
\newcommand{\chipp}{{\chi^{\prime \prime}}}
\newcommand{\chippp}{{\chi^{\prime \prime \prime}}}

\renewcommand{\Re}{\operatorname{Re}}
\renewcommand{\Im}{\operatorname{Im}}

\newcommand\GReq{\mathrel{\overset{\makebox[0pt]{\mbox{\normalfont\tiny\sffamily GR}}}{=}}}

\def\n{\noindent}


\newcommand{\Sussex}{Department of Physics \& Astronomy, University of Sussex, Brighton BN1 9QH, UK}


\title{Effect of lensing non-Gaussianity on the CMB power spectra}
\author{Antony Lewis}
\affiliation{\Sussex}
\author{Geraint Pratten}
\affiliation{\Sussex}
\homepage{http://cosmologist.info}

\begin{abstract}

Observed CMB anisotropies are lensed, and the lensed power spectra can be calculated accurately
assuming the lensing deflections are Gaussian. However, the lensing deflections are actually slightly non-Gaussian due to
both non-linear large-scale structure growth and post-Born corrections. We calculate the leading correction to the lensed CMB power spectra from the non-Gaussianity, which is determined by the lensing bispectrum.
Assuming no primordial non-Gaussianity, the lowest-order result gives $\sim 0.3\%$ corrections to the BB and EE polarization spectra on small-scales. However we show that the effect on EE is reduced by about a factor of two by higher-order Gaussian lensing smoothing, rendering the total effect safely negligible for the foreseeable future. We give a simple analytic model for the signal expected from skewness of the large-scale lensing field; the effect is similar to a net demagnification and hence a small change in acoustic scale (and therefore out of phase with the dominant lensing smoothing that predominantly affects the peaks and troughs of the power spectrum).
\end{abstract}

\pacs{
}

\maketitle

\begin{acronym}
\acrodef{WL}[WL]{Weak Lensing}
\end{acronym}

\newcommand{\WL}{\ac{WL}\xspace}

\section{Introduction}
\label{sec:intro}

Lensing must be accurately modelled to understand the observed CMB power spectra, as it changes the power spectra at the $5$--$30\%$ level on small scales and also introduces B modes~\cite{Seljak:1996ve,Hu:2000ee,Challinor:2005jy} (see Ref.~\cite{Lewis:2006fu} for a review).
In a previous paper, Ref.~\cite[hereafter PL16]{Pratten:2016dsm}, we calculated the leading order bispectrum
of the lensing deflections. The bispectrum arises from changes to the lensing convergence from non-linear large-scale structure (LSS) growth, and non-linear convergence and rotation from post-Born corrections. The non-Gaussianity for the CMB lensing deflections is small, since the non-linear structure growth signal is substantially suppressed by the large number of lenses along the line of sight, and the lensing kernel peaks at quite high redshifts where the gravitational potentials are still nearly linear. However, the bispectrum will still easily be detectable, and coincidentally is the same order of magnitude as the bispectrum from post-Born lensing (which grows with source-plane distance); both effects need to be included, and for non-equilateral shapes partly cancel.

In this paper we give the first full calculation of the lensed CMB power spectra including the leading correction from non-Gaussianity of the lensing deflection field. In \paperone\ we calculated the effect of the purely post-Born convergence-convergence-rotation bispectrum on the lensed B modes; we now extend this to also include the full convergence bispectrum from LSS and post-Born terms, and assess the impact of all the temperature and polarization power spectra. Ref.~\cite{Marozzi:2016uob} have calculated the leading effect of the post-Born bispectrum on the lensed CMB temperature, but since they did not include the LSS contribution their result is not directly observable (and misestimates the sign of the full effect since the LSS terms have opposite sign and are somewhat larger). By working to leading order in the non-Gaussian signal, but non-perturbatively in the Gaussian lensing, we shall also
see that higher-order Gaussian lensing significantly reduces the total predicted bispectrum contribution compared to that predicted by the lowest-order fully-perturbative result.

Our notation and conventions follow \paperone, so for brevity we do not repeat most of the introduction or detail of the bispectrum calculation here. We start in Sec.~\ref{sec:perturbative} by considering a series expansion in the curl and gradient deflection angles, and derive the leading perturbative effect on the lensed CMB power spectra. The results only depend on the total bispectrum, which can include post-Born and LSS terms. In Sec.~\ref{sec:analytic} we give a simple approximate analytic model that describes qualitatively the main effect and gives some intuition about where the signal is coming from.
Finally in Sec.~\ref{sec:nonpert} we generalize to our final result for the leading non-Gaussian correction, including non-perturbatively (but approximately) higher order Gaussian lensing smoothing effects that have a significant impact on the total size of the signal. This then allows us to assess the observational significance, and hence whether it is safe to neglect the bispectrum when calculating CMB power spectra.

\section{Leading lensed power spectra from deflection non-Gaussianity}
\label{sec:perturbative}
We are interested in the effect of lensing on power spectra of the CMB temperature (T), and polarization (E and B) fields. The lensing deflection angle may be decomposed into two lensing potentials
\begin{align}
\alpha_a &= \nabla_a \phi + \epsilon_{ab} \nabla^b \Omega ,
\end{align}
\n
where $\phi$ is the lensing potential (describing convergence $\kappa = -\vgrad^2\phi/2$), $\Omega$ is the curl potential (describing rotation $\omega = -\vgrad^2\Omega/2$), and $\nabla$ is the covariant angular derivative. Since the few-arcminute deflections can become comparable to the scale of the CMB modes being lensed, the lensed power spectra are most accurately calculated using a non-perturbative approach following Refs.~\cite{Seljak:1996ve,Challinor:2005jy} in the approximation in which the lensing convergence is Gaussian and uncorrelated to the CMB.  Equivalent results for lensing by rotation in the Gaussian approximation are given in Ref.~\cite{Padmanabhan:2013xfa}. Non-linear corrections to the convergence power spectrum are dominated by non-linear LSS growth, and are easily incorporated given a fitting function for the non-linear matter power spectrum, though this may have significant modelling uncertainties on small scales. As shown in~\paperone\ corrections to the convergence power spectrum from post-Born lensing are negligible. Post-Born lensing does however introduce a qualitatively new rotation signal, and the corresponding small effects on the power spectra in the Gaussian approximation are calculated in detail in \paperone.
Correlations between the CMB and lensing potential are only significant on large scales and can be neglected to high accuracy (though correlations affecting foreground contaminants could be much more significant); we neglect them here and focus on the corrections from relaxing the Gaussian assumption.

The lensed CMB power spectra can be defined as
\be
\la \tilde{X}_{\ell m} \tilde{Y}_{\ell m}^*\ra = \left.\tilde{C}_{\ell}^{XY}\right|_G + \Delta \tilde{C}_{\ell}^{XY},
\ee
where $ \left.\tilde{C}_{\ell}^{XY}\right|_G$ is the non-perturbative result for the lensed CMB power spectra in the Gaussian approximation and ${\tilde{X},\tilde{Y}} \in {\tilde{T}, \tilde{E},\tilde{B}}$ are the observed lensed CMB fields.  The correction $\Delta C_{\ell}^{XY}$ due to non-Gaussianity of the lensing fields is small, and does not need to be calculated to high precision. In this paper we calculate it to leading order in the perturbations, which is $\clo(\Psi^4)$, where $\Psi$ is the gravitational Weyl potential. Since the small correction is only likely to be significant compared to cosmic variance on small scales, we calculate it using the flat-sky approximation (the leading Gaussian term can of course be calculated using the more accurate full-sky result of Refs.~\cite{Challinor:2005jy,Lewis:2006fu}).

 Working in the flat sky approximation, the lensed temperature anisotropies can be written as
\begin{align}
\tilde{T} (\bft) &= \int \frac{d^2 \bfell}{(2 \pi)^2} \tilde{T}({\bfell}) \, e^{i \bfell \cdot \bft} .
\end{align}
\n
The lensed polarization anisotropies can be defined in an analogous way in terms of the spin $\pm2$ Stokes parameters $\tilde{Q} \pm i \tilde{U}$ and the more physically relevant $\tilde{E}$ and $\tilde{B}$ modes
\begin{align}
\left[ \tilde{Q} \pm i \tilde{U} \right] (\bft) &= - \int \frac{d^2 \bfell}{(2\pi)^2} \, \left( \tilde{E}({\bfell}) \pm i \tilde{B}({\bfell}) \right) \, e^{\pm i 2 \varphi_{\ell}} \, e^{i \bfell \cdot \bft} .
\end{align}
\n
The unlensed CMB observables $\lbrace T_{\ell},E_{\ell},B_{\ell} \rbrace$ are defined in exactly the same way. The lensed (or unlensed) angular power spectra in the flat-sky approximation are then
\begin{align}
\la \tilde{X}({\bfell})\tilde{Y}({\bfell'}) \ra &= (2 \pi)^2 \delta_D(\bfell + \bfell') \, \tilde{C}^{XY}_{\ell} .
\end{align}

We expand a lensed spin-s field $\tilde{\zeta}^s(\bft)$ as a perturbative expansion in terms of the unlensed field $\zeta^{s}$ as\footnote{
Additional small corrections due to polarization rotation, time delay, and the lensed emission angle not being orthogonal to the background last-scattering surface are calculated in Ref.~\cite{Lewis:2017ans}.}
 \cite{Hu:2000ee,Lewis:2006fu}
\begin{align}
\tilde{\zeta}^s (\bft) &= \zeta^s (\bft) + \valpha_a \nabla^a \, \zeta^s (\bft) + \frac{1}{2} \valpha_a \valpha_b \nabla^a \nabla^b \, \zeta^s (\bft) + \frac{1}{6} \valpha_a \valpha_b \valpha_c \nabla^a \nabla^b \nabla^c \, \zeta^s (\bft) + \mathcal{O}(\alpha^4).
\label{eq:alphaseries}
\end{align}
Note that the trispectrum $\la \alpha_a \alpha_b \alpha_c \alpha_e\ra$ has a disconnected Gaussian piece which is already $\clo(\Psi^4)$ from the linear deflection angles, hence any non-linear contribution to the connected trispectrum is of higher order: to lowest order we can consistently neglect all N-point functions above the bispectrum. The $\alpha^3$ term in Eq.~\eqref{eq:alphaseries} could give an $\clo(1)\times \clo(\alpha^3)$ bispectrum contribution to the lensed power spectra; however
$\la  \valpha_a \valpha_b \valpha_c\ra =0$ since there is no symmetric isotropic three tensor, and this term vanishes. Expanding the remaining relevant terms into harmonics we have the series approximation
\begin{multline}
\tilde{\zeta}^s(\bfell) \approx  {\zeta}^s(\bfell) - \int \frac{d^2 \bfell_1}{(2 \pi)^2} \, \zeta^s (\bfell_1) \, e^{i s \varphi_{\bfell_1 \bfell}} \, \left[ \bfell_1 \cdot \left( \bfell - \bfell_1 \right) \phi(\bfell - \bfell_1) + \bfell_1 \times \bfell \,\Omega (\bfell - \bfell_1) \right] \\
 -\frac{1}{2} \int \frac{d^2 \bfell_1}{(2 \pi)^2} \int \frac{d^2 \bfell_2}{(2 \pi)^2} \, \zeta^s (\bfell_1) \, e^{i s \varphi_{\bfell_1 \bfell}}  \,\biggl( \left[ \bfell_1 \cdot \bfell_2 \, \phi(\bfell_2) + \bfell_1 \times \bfell_2 \, \Omega(\bfell_2) \right] \\
\times \left[ \bfell_1 \cdot \left( \bfell_2 + \bfell_1 - \bfell \right) \, \phi(\bfell - \bfell_1 - \bfell_2) + \bfell_1 \times (\bfell_2 - \bfell) \, \Omega(\bfell - \bfell_1 - \bfell_2) \right]
\label{eqn:expansion}
\end{multline}
where the cross product is defined by $\bfell \times \bfL \equiv \epsilon_{ab} \ell^a L^b$, and $\varphi_{\bfell_1 \bfell} = \varphi_{\bfell_1} - \varphi_{\bfell}$ is the angle between $\bfell_1$ and $\bfell$.

Bispectra arise from correlation of three lensing potential fields, and are defined under statistical isotropy on the flat-sky by
\begin{align}
\langle X (\vL_1) Y (\vL_2) Z (\vL_3) \rangle &= (2 \pi)^2 \delta_D (\vL_1 + \vL_2 + \vL_3 ) \, b_{L_1,L_2,L_3}^{XYZ}.
\end{align}
The curl potential $\Omega$ is only generated at second order from post-Born effects, so to $\clo(\Psi^4$) we only need to consider the two bispectra $b^{\kappa\kappa\kappa}_{\ell_1, \ell_2,\ell_3} = \frac{\ell_1^2\ell_2^2\ell_3^2}{8} b^{\phi\phi\phi}_{\ell_1, \ell_2,\ell_3}$ and  $b^{\kappa\kappa\omega}_{\ell_1, \ell_2,\ell_3} = \frac{\ell_1^2\ell_2^2\ell_3^2}{8} b^{\phi\phi\Omega}_{\ell_1, \ell_2,\ell_3}$ at leading order.
The $\kappa\kappa\omega$ bispectrum only has post-Born contributions at leading order, but the convergence bispectrum has contributions from both post-Born and LSS non-linearities, where explicit expressions and calculational details are given in \paperone. We assume negligible primordial non-Gaussianity.

We can now proceed to calculate the bispectrum contribution to the lensed power spectra.
Using Eq.~\eqref{eqn:expansion} the $\clo(\phi)\times \clo(\phi^2)$ contribution from the convergence bispectrum assuming no unlensed B modes is
\begin{multline}
\Delta\tilde{C}^{XY}_{\ell}[\text{convergence}] = - \int \frac{d^2 \bfell_1}{(2 \pi)^2}  \int \frac{d^2 \bfell_2}{(2 \pi)^2}
\trig_{XY}(\varphi_{\bfell_1 \bfell})
\left[ \bfell_1 \cdot (\bfell + \bfell_1) \right] \left[ \bfell_1 \cdot \bfell_2 \right] \left[ \bfell_1 \cdot (\bfell_1 + \bfell_2 + \bfell ) \right]
\\
\times  C^{\bar{X}\bar{Y}}_{\ell_1} b^{\phi\phi\phi} (\bfell + \bfell_1 , \bfell_2 , -\bfell - \bfell_1 - \bfell_2) ,
\label{eq:convergence}
\end{multline}
where $\trig_{TT}(\varphi)=1$, $\trig_{EE}(\varphi)=\cos^2(2\varphi)$, $\trig_{BB}(\varphi)=\sin^2(2\varphi)$, $\trig_{TE}(\varphi)=\cos(2\varphi)$, mixed parity terms are zero and $\bar{T}=T$, $\bar{E}=E$ and $\bar{B}=E$.

Contributions involving the curl potential are a $\clo(\Omega)\times \clo(\phi^2)$ (``correlation'')
contribution that correlates the leading rotation contribution with the second-order potential term
\begin{multline}
\Delta\tilde{C}^{XY}_{\ell}[\text{correlation}] =
 \int \frac{d^2 \bfell_1}{(2 \pi)^2}  \int \frac{d^2 \bfell_2}{(2 \pi)^2}
\trig_{XY}(\varphi_{\bfell_1 \bfell}) \left[ \bfell_1 \cdot \bfell_2 \right]  \left[ \bfell \times \bfell_1 \right] \left[ \bfell_1 \cdot (\bfell_2 + \bfell_1 + \bfell ) \right] \\
\times C^{\bar{X}\bar{Y}}_{\ell_1} b^{\phi \phi \Omega} (\bfell_2 , - \bfell - \bfell_1 - \bfell_2 , \bfell + \bfell_1 ),
\label{eq:correlation}
\end{multline}
and a $\clo(\phi)\times \clo(\phi\Omega)$ (``mixed'') contribution:
\begin{multline}
\Delta\tilde{C}^{XY}_{\ell}[\text{mixed}] =
- 2\int \frac{d^2 \bfell_1}{(2 \pi)^2}  \int \frac{d^2 \bfell_2}{(2 \pi)^2}\trig_{XY}(\varphi_{\bfell_1 \bfell})
\left[ \bfell_1 \cdot \bfell_2 \right] \,
\left[ \bfell_1 \cdot (\bfell + \bfell_1 ) \right] \,  \left[ \bfell_1 \times ( \bfell_2 + \bfell ) \right]
\\ \times C^{\bar{X}\bar{Y}}_{\ell_1}  b^{\phi \phi \Omega} (\bfell + \bfell_1 , \bfell_2 , - \bfell - \bfell_1 - \bfell_2).
\label{eq:mixed}
\end{multline}

\begin{figure*}
\begin{center}
\includegraphics[width=180mm]{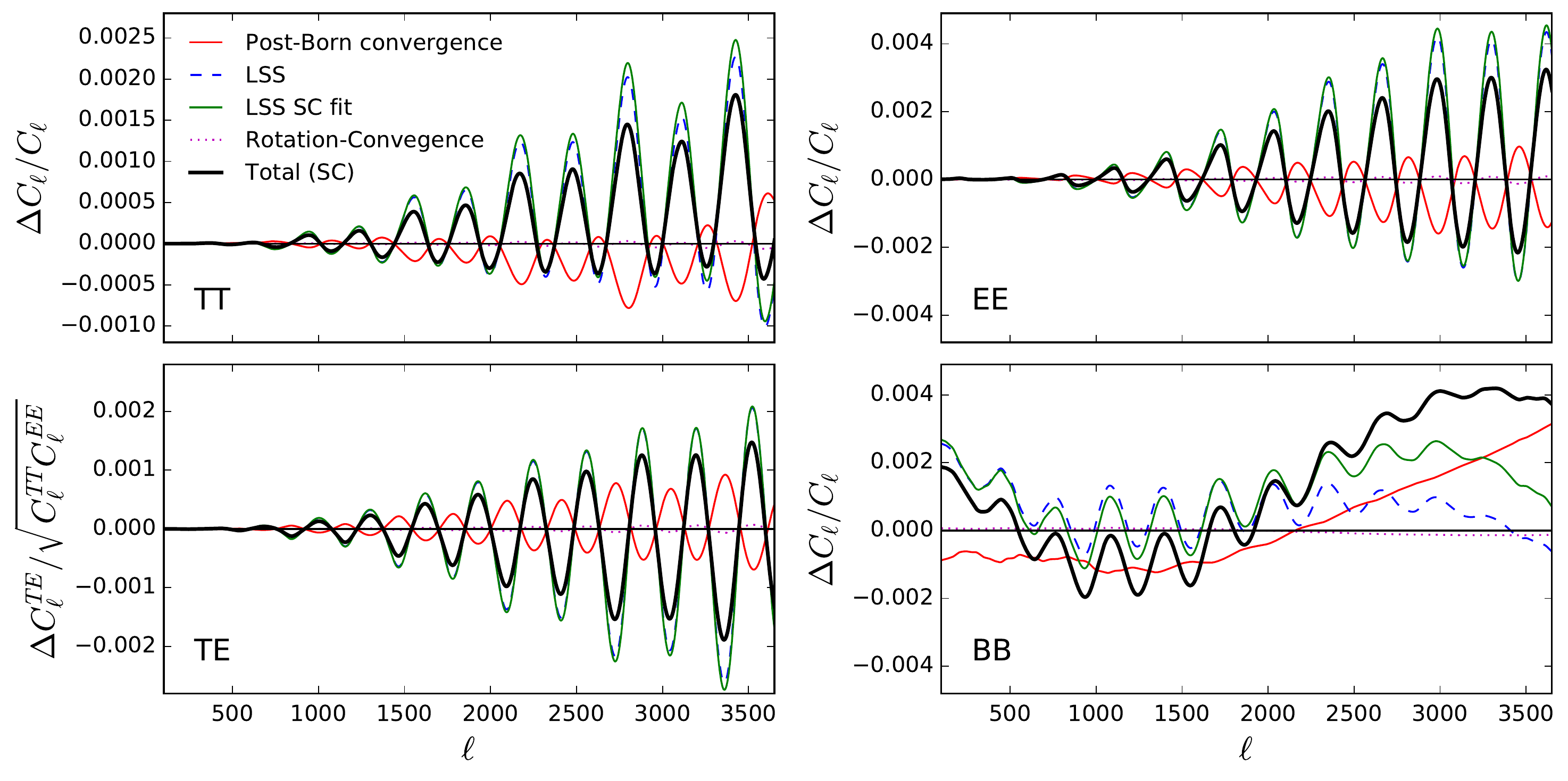}
\caption{Fractional change in the lensed CMB power spectra due to the leading correction from the lensing bispectrum (and neglecting higher-order Gaussian lensing on the correction).
Contributions from the post-Born convergence bispectrum (red) and LSS bispectrum are mostly of opposite sign and partly cancel (tree-level LSS result is shown dashed, non-linear fitting formula of \citet{Scoccimarro:2000ee} (``SC'') is shown in green). The dotted line shown the very small contributions from the two $\kappa\kappa\omega$ post-Born bispectrum terms. The thick black line shows the total effect.
}
\label{fig:BispectrumCls}
\end{center}
\end{figure*}

\begin{figure*}
\begin{center}
\includegraphics[width=180mm]{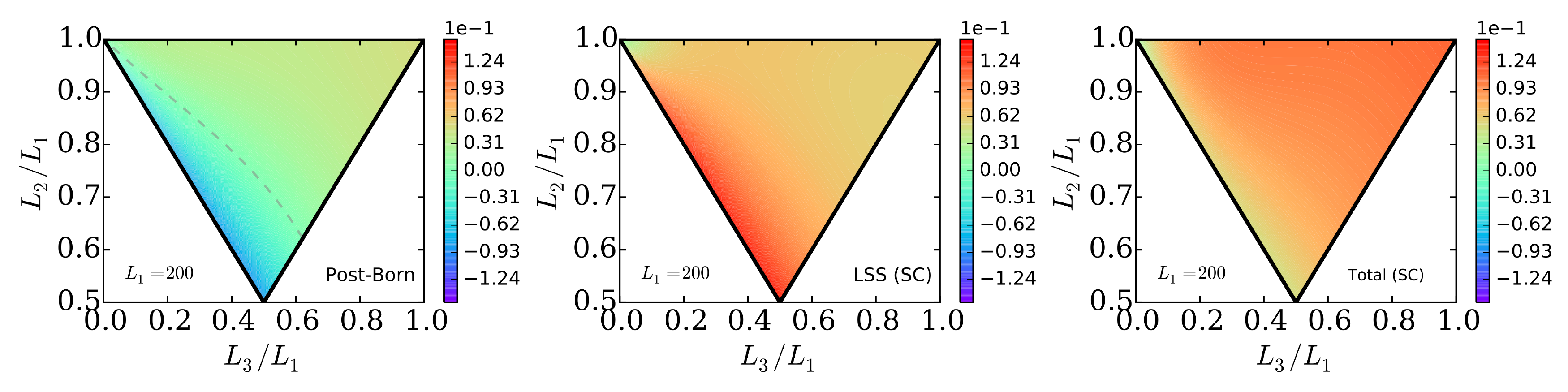}
\caption{
{
Slices through the weighted convergence bispectrum $(L_2 L_3)^{1/2} \, b^{\kappa\kappa\kappa}_{L_1 L_2 L_3} / (C^{\kappa\kappa}_{L_1} C^{\kappa\kappa}_{L_2} C^{\kappa\kappa}_{L_3} )^{1/2}$ for $L_1 = 200$. The left plot shows post-Born contribution to the bispectrum, the middle plot the LSS bispectrum using the non-linear fit of \citet{Scoccimarro:2000ee} (``SC'') and the right plot the total bispectrum. The grey dashed line denotes the $b^{\kappa\kappa\kappa}=0$ contour. For the T and E spectra, the bulk of the signal comes from the bispectrum with lensing modes on relatively large scales $L \alt 250$ where the difference between the tree-level and perturbative bispectrum is relatively small.
}
}
\label{fig:bispectrum_kkk_PB_LSS_triangle}
\end{center}
\end{figure*}

Numerical results for the fractional change in the lensed CMB power spectrum due to the bispectrum contributions are shown in Fig.~\ref{fig:BispectrumCls}. Our result for the change in the TT spectrum due to the post-Born convergence bispectrum agrees well with Ref.~\cite{Marozzi:2016uob}. However, both post-Born and LSS convergence bispectra contribute at similar levels and mostly with opposite sign; in fact the LSS contribution is somewhat larger, and hence determines the sign of the total result. The fact that the LSS and post-Born terms partly cancel can be understood in terms of the relative sign of the bispectra for flattened triangle shapes as explained in detail in \paperone. The cancellation is illustrated at the bispectrum level in Fig.~\ref{fig:bispectrum_kkk_PB_LSS_triangle}, which shows a slice through the bispectra for relatively large scales ($L\alt 200$) relevant for the lensing smoothing effect on the power spectra:
for flattened shapes the post-Born bispectrum is negative, and partly cancels the larger LSS signal there. For near-equilateral shapes both contributions are positive, but these do not have a large effect on the lensed power spectra. The convergence bispectrum effect is dominated by near-flattened triangles because a CMB mode is only lensed if the deflection angles have component in the direction of change of the unlensed CMB; mathematically, all three arguments of the bispectrum in Eq.~\eqref{eq:convergence} are dotted into $\bfell_1$, the unlensed CMB mode.

The lensing modes responsible for the lensing of the power spectra are relatively large scale, and hence the LSS bispectrum is expected to be well approximated by the tree-level result of Ref.~\cite{Bernardeau:2001qr} (evaluated using fully non-linear power spectra from \HALOFIT\ \cite{Smith:2002dz,Takahashi:2012em} for better accuracy).
To see the typical size of effects beyond tree level we also compare to results using the fitting formula of ~\citet{Scoccimarro:2000ee} (``SC''). For the T and E spectra the lensing smoothing, and hence bulk of the signal comes from the bispectrum, comes from lensing modes that are all relatively large scale ($L\alt 250$), so the perturbative tree-level result is quite accurate. The BB spectrum is more sensitive to smaller scale power, and hence Fig.~\ref{fig:BispectrumCls} shows a larger difference between the tree-level and fitting formula results for BB at high $\ell$. Our bispectrum results are calculated in the lowest order Limber approximation; since the CMB power spectrum lensing effect is somewhat sensitive to $L\alt 40$ where this approximation becomes inaccurate, this could lead to a small error. However, since the bispectrum correction we calculate is small, the error on the correction should be negligible.

As shown in Fig.~\ref{fig:BispectrumCls} the rotation bispectrum contributions from Eqs.~\eqref{eq:correlation}
and \eqref{eq:mixed} are an order of magnitude smaller, at most the $\clo(10^{-4})$ level, and hence negligible. The correlation term of Eq.~\eqref{eq:correlation} gives a qualitatively new correlation between the leading rotation-induced lensed B modes and the convergence-induced B modes (see \paperone\ for further detailed discussion), but the small size of the correlations means that the rotation and convergence induced B modes can still be regarded as uncorrelated to a good approximation.

The full result is well approximated by just the sum of the LSS and post-Born convergence terms, since the rotation contributions are negligible. At $\ell < 2000$ the total effect is still below cosmic variance and hence clearly negligible. The fractional effect on the B-mode signal grows to $\sim 0.4\%$ on small scales; however, the B-mode power spectrum is itself small, has strong off-diagonal correlations, and is sensitive to small-scale lenses (introducing considerable non-linear modelling uncertainties from modelling the small-scale matter power spectrum). The B-mode signal is therefore likely to be negligible in practice. On smaller scales the EE signal also grows, being nearly $0.3\%$ in EE at $\ell \agt 3000$, which looks as though it could be clearly observationally relevant. However, this was just a lowest order calculation, and we will need to consider higher-order Gaussian lensing effects to get a more reliable estimate. But first we discuss a simple analytic model to develop some intuition for the signal being generated.

\subsection{Analytic model}
\label{sec:analytic}

\begin{figure*}
\begin{center}
\includegraphics[width=120mm]{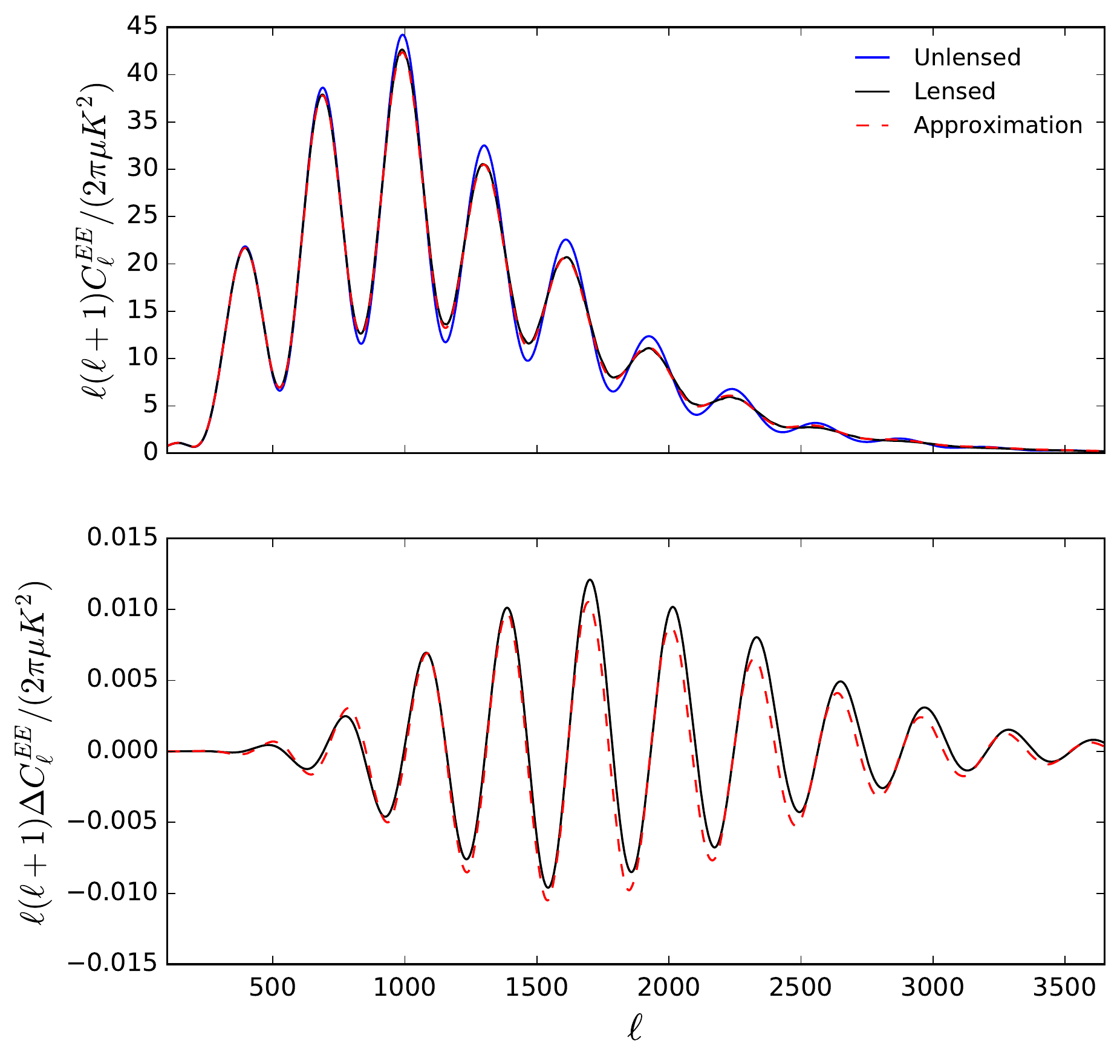}
\caption{
\emph{Top}: The Gaussian lensing effect on the EE power spectrum, compared to the squeezed analytic model of Eq.~\eqref{eq:analytic} (taking $\la\kappa^2\ra = 5\times 10^{-4}$, corresponding approximately to the convergence power at $L<250$). The analytic model agrees well by eye, though the fractional error becomes quite large on very small scales. \emph{Bottom}: the change in the EE power spectrum due to the convergence bispectrum (Eq.~\ref{eq:convergence}), compared to the analytic model of Eq.~\eqref{eq:analytic} with skewness parameter $\la \kappa^3\ra = 2.5\times 10^{-7}$. The values of $\la\kappa^2\ra$ and $\la \kappa^3\ra$ in the analytic models are chosen by hand to approximately fit the numerical amplitude.
}
\label{fig:squeezed}
\end{center}
\end{figure*}

The lensing modes responsible for the bulk of the lensing of T and E are relatively large scale (most of the bispectrum signal comes from modes $L<250$). To consider the local impact on the small-scale modes, it is useful to approximate the large scale lensing field by locally constant convergence and shear values.
For simplicity consider just a convergence\footnote{We give the more general result including shear in Appendix~\ref{app:squeezed}}, so the locally constant magnification matrix is $A_{ij} = (1-\kappa)\delta_{ij}$. For a local displacement $\zeta$ we have lensed field $\tilde{X} = X(\mA \zeta)$, so taking the
expectation of the unlensed small-scale CMB with fixed $\mA$ we have~\cite{Bucher:2010iv,Lewis:2011fk,Bonvin:2015uha}
\be
\la \tilde{X}(\bfell)\tilde{Y}(\bfell')\ra_{X,Y} = C^{XY}_{|\mA^{-1}\bfell|} \frac{\delta(\bfell+\bfell')}{|\mA|}.
\ee
Expanding in powers of the fixed $\kappa$ we then have
\be
\left.\ell^2\tilde{C}_\ell^{XY}\right|_{\kappa} = \ell^2|\mA|^{-1} C^{XY}_{|\mA^{-1}\bfell|} =
\ell^2 C^{XY}_{\ell} +
\kappa \frac{\ud \ell^2C_\ell^{XY}}{d\ln \ell}
+\frac{1}{2}\kappa^2 \left( \frac{\ud (\ell^2C_\ell^{XY})}{d\ln \ell} + \frac{\ud^2 (\ell^2C_\ell^{XY})}{\ud\ln\ell^2} \right)
+ \frac{1}{6}\kappa^3 \ell\frac{ \ud^3 (\ell^4C_\ell^{XY})}{\ud \ell^3} +\dots
\ee
The first order term describes the shift in angular scale due to the local $\kappa$ value (which is used for lensing reconstruction). Averaging over $\kappa$ values this term vanishes, and we have the lensed spectrum
\be
\ell^2\tilde{C}_\ell^{XY} =
\ell^2 C^{XY}_{\ell}
+\frac{1}{2}\la\kappa^2\ra \left( \frac{\ud (\ell^2C_\ell^{XY})}{d\ln \ell} + \frac{\ud^2 (\ell^2C_\ell^{XY})}{\ud\ln\ell^2} \right)
+ \frac{1}{6}\la\kappa^3\ra \ell\frac{ \ud^3 (\ell^4C_\ell^{XY})}{\ud \ell^3} +\dots
\label{eq:analytic}
\ee
 The variance term describes the lensing smoothing effect on the peaks and troughs.
 The third term describes the effect of skewness, which is out of phase with the smoothing as it depends on the third rather than second derivative of the spectrum.
 As shown in Fig.~\ref{fig:squeezed} this approximation actually does a good job of describing the main effects on the T and E mode spectra (after choosing the variance and skewness parameters appropriately). The third derivative describes a small sideways shift of the acoustic peaks:
 non-linear clumping of matter gives concentrated areas of high density and convergence (like filaments), surrounded by larger areas of below-average density and convergence, corresponding to a positive bispectrum and skewness (e.g. see Ref.~\cite{Lewis:2011au}). The larger area of low density gives an effective net demagnification of the CMB acoustic scale, corresponding to a small average shift of the small-scale acoustic peaks to higher $\ell$.

\subsection{Impact of higher-order lensing corrections}
\label{sec:nonpert}
The result of Eq.~\eqref{eq:convergence} is based on the series expansion of the lensed field about a totally unlensed field. However since Gaussian lensing changes the CMB power spectra by $\clo(1)$ (by an amount comparable to the size of the unlensed signal) on small scales, the result may be biased due to the neglect of higher order terms. Any three lensing modes that modify the CMB power due to a non-zero bispectrum will in practice modify the CMB affected by other lensing modes, not the entirely unlensed CMB: the fully unlensed $C_\ell^{XY}$ that appears in Eq.~\eqref{eq:convergence} may not be the right quantity at the $\clo(1)$ level.

To model the non-perturbative effects, we can follow a similar method to that used for getting the correct non-perturbative weights for lensing reconstruction and the bispectrum following Ref.~\cite{Lewis:2011fk}. Since we do not require the result to high accuracy, we can note that for small-scale CMB modes and large scale lenses we can approximate the geometric factors in Eq.~\eqref{eq:convergence} that distinguish the T and E as being the same ($\trig_{XY}\approx \cos(0)\approx 1$, since $\vell \approx -\vell_1$ to couple to high-$\ell$ CMB to large-scale lenses). The formulae for lensing T and E will therefore be approximately equivalent, so we only need to consider the simplest temperature case (we do not consider the B-modes spectrum here, since the perturbative bispectrum contributions are already very small and the calculation would be more complicated since the kernel is not approximately the same as for T and E). Also, since the bispectrum term is dominated by the convergence bispectrum, we neglect the rotation.

Consider the leading response of two small-scale CMB modes to three convergence lensing modes with non-zero bispectrum
\begin{align}
\la \tilde{X}(\vell)\tilde{Y}(\vell')\ra = \la \tilde{X}(\vell)\tilde{Y}(\vell')\ra_G +
\frac{1}{6}\int \ud^2\vell_1 \ud^2\vell_2 \ud^2\vell_3 \left\la \frac{\delta \left(\tilde{X}(\vell)\tilde{Y}(\vell')\right)}{\delta \phi(\vell_1)\delta\phi(\vell_2)\delta\phi(\vell_3)} \right\ra_G
\la \phi(\vell_1)\phi(\vell_2)\phi(\vell_3)\ra + \dots
\label{bresponse}
\end{align}
In terms of the power spectra and bispectrum this is
\begin{align}
\delta_D(\vell+\vell') \tilde{C_\ell^{XY}}= \delta_D(\vell+\vell')\left. \tilde{C}^{XY}_\ell\right|_G +
\frac{1}{6}\int \ud^2\vell_1\ud^2\vell_2 \left\la \frac{\delta \left(\tilde{X}(\vell)\tilde{Y}(\vell')\right)}{\delta \phi(\vell_1)\delta\phi(\vell_2)\delta\phi(\vell_3)} \right\ra_G
b^{\phi\phi\phi}_{\ell_1\ell_2\ell_3}+ \dots
\end{align}
Using $\frac{\delta}{\delta\phi(\vell)} = \frac{\delta\valpha_a}{\delta\phi(\vell)}\grad^a$ we have
\be
\frac{\delta \tilde{X}(\vell)}{\delta\phi(\vell_1)} = \frac{i}{(2\pi)^2}\ell_1^a
\widetilde{\grad_a\! X}(\vell-\vell_1),
\qquad
\frac{\delta \tilde{X}(\vell)}{\delta\phi(\vell_2)\delta\phi(\vell_3)}
= -\frac{1}{(2\pi)^4}\ell_2^a\ell_3^b \widetilde{\grad_a\!\grad_b\! X}(\vell-\vell_2-\vell_3),
\ee
where the tilde denotes the lensed quantity (i.e. the Fourier transform of the gradients at lensed positions $\vx + \valpha$).
The third derivative term does not contribute as it vanishes after angular integration in Eq.~\eqref{bresponse}. In Appendix~\ref{Cgrads} we show that
\begin{align}
\left\la \widetilde{\grad_a \grad_b X}(\vell) \widetilde{\grad_c Y}(\vell') \right\ra
    = i(2\pi)^2 \delta(\vell+\vell')\left(\tilde{C}^{\eth^2X\eth Y}_\ell \ell_{\la a}\ell_b\ell_{c\ra} + \frac{3\ell^2}{4} \tilde{C}^{\grad^2 X\grad Y}_\ell \ell_{(a}\delta_{bc)}  \right) \approx
    i(2\pi)^2 \delta(\vell+\vell')\tilde{C}^{\eth^2X\eth Y}_\ell \ell_a\ell_b\ell_c
    \label{eq:gradpowerapprox}
\end{align}
and give explicit forms for the two lensed gradient power spectra for numerical calculation. These new spectra are similar to the lensed power spectrum, but with the lensing smoothing applied to the power spectrum multiplied by $\ell^3$; numerically the two power spectra are quite similar (see Fig.~\ref{fig:lensedspectra}), and at our level of approximation can take them to be the same, hence the final approximation in Eq.~\eqref{eq:gradpowerapprox}. In the limit of zero lensing the gradient spectra are defined so that $\tilde{C}^{\eth^2X\eth Y}_\ell\rightarrow C_{\ell}^{XY}$.
\begin{figure*}
\begin{center}
\includegraphics[width=160mm]{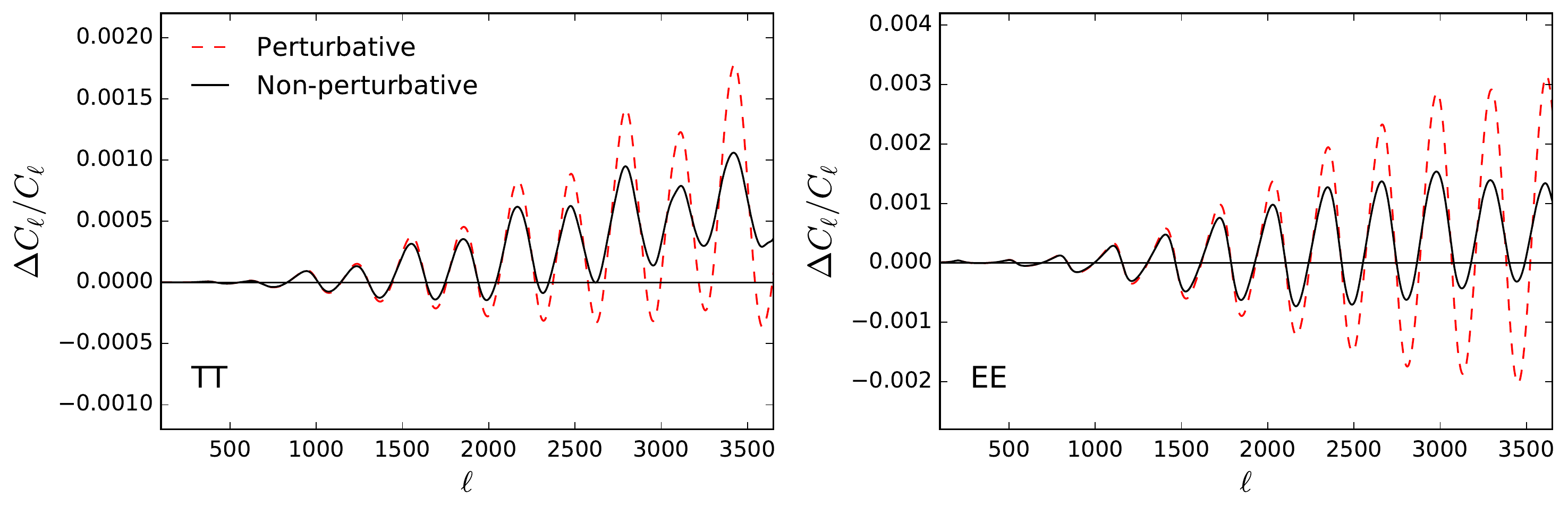}
\caption{
The fractional change in the lensed temperature and E-mode polarization CMB power spectra due to the full post-Born plus LSS (SC) convergence bispectrum, using the result to lowest order in the Gaussian lensing of Eq.~\eqref{eq:convergence} (dashed) compared to the more accurate non-perturbative result of Eq.~\eqref{eq:convergence_nonpert} (solid).
The bispectrum response approximately goes like a derivative of the gradient spectrum, and hence is significantly smaller using the lensing-smoothed non-perturbative result on small scales. The final $\alt 0.15\%$ change is negligible, and the cross-correlation TE spectrum behaves similarly.
}
\label{fig:smoothed}
\end{center}
\end{figure*}

Using this result we have
\be
\left\la \frac{\delta \left(\tilde{X}(\vell)\tilde{Y}(\vell')\right)}{\delta \phi(\vell_1)\delta\phi(\vell_2)\delta\phi(\vell_3)} \right\ra_G
\approx \frac{6}{(2\pi)^4} [(\vell+\vell_1)\cdot \vell_1][(\vell+\vell_1)\cdot \vell_2][(\vell+\vell_1)\cdot \vell_3] \delta(\vell+\vell')\tilde{C}^{\eth^2X\eth Y}_{|\vell+\vell_1|}.
\ee
Hence after a change of integration variable our more accurate result for T and E spectra, which is non-perturbative in the Gaussian lenses, is finally
\begin{multline}
\Delta\tilde{C}^{XY}_{\ell}[\text{convergence}] \approx  \int \frac{d^2 \bfell_1}{(2 \pi)^2}  \int \frac{d^2 \bfell_2}{(2 \pi)^2}
\trig_{XY}(\varphi_{\bfell_1 \bfell})\left[ \bfell_1 \cdot (\bfell + \bfell_1) \right] \left[ \bfell_1 \cdot \bfell_2 \right] \left[ \bfell_1 \cdot (\bfell_1 + \bfell_2 + \bfell ) \right]
\\
\times  \tilde{C}^{\eth^2X\eth Y}_{\ell_1} b^{\phi\phi\phi} (\bfell + \bfell_1 , \bfell_2 , -\bfell - \bfell_1 - \bfell_2) .
\label{eq:convergence_nonpert}
\end{multline}
Here we manually re-inserted the $\trig_{XY}(\varphi_{\bfell_1 \bfell})\approx 1$ factor for consistency with the weak lensing limit in the case of polarization.
This result is then the same form as Eq.~\eqref{eq:convergence} but using the lensed gradient power spectrum $\tilde{C}^{\eth^2X\eth Y}$ instead of the unlensed spectrum. Since the lensed gradient power spectra are smoother than the unlensed spectra, and since (as discussed in Sec.~\ref{sec:analytic}) the response to the bispectrum goes like a third derivative of the spectrum, the net effect is that the amplitude of the bispectrum term is significantly reduced on small scales. Fig.~\ref{fig:smoothed} shows how the $\Delta \tilde{C}_\ell$ from the bispectrum is reduced in amplitude from nearly $\sim 0.3\%$ on small scales for EE, to at most $\approx 0.15\%$. The final effect on the EE signal is detectable at less than $2\sigma$ even with perfect cosmic-variance limited noise-free data to $\ell\sim 4000$, and hence negligible. The temperature signal is somewhat smaller, and the cross-correlation follows a similar pattern.


\section{Conclusions}

We presented new analytic results for the leading corrections to the lensed CMB power spectra from the lensing bispectrum, including non-perturbative Gaussian smoothing which is quantitatively important for the overall size of the signal. Our final calculation of the change in the power spectra due to the lensing bispectrum contribution is safely negligible at $\ell < 4000$. Since the temperature and E-polarization signals are dominated by large-scale lenses where the tree-level bispectrum result is quite accurate, this conclusion should be robust to uncertainties in the modelling of non-linear LSS bispectrum. For B-mode polarization smaller-scale lenses are more important, and hence more strongly non-linear non-Gaussianity could give larger corrections.
Results from ray tracing numerical simulations, e.g. following Ref.~\cite{Calabrese:2014gla}, should of course include both the LSS and post-Born effects non-perturbatively, but the analytic results should be sufficiently accurate to model of the signal if required to avoid small biases when analysing far-future small-scale polarization data. With high signal to noise measurements of the large-scale lensing modes the observed sky can also be delensed at the map level, which would reduce the bispectrum signal further.

In this paper we assumed no primordial non-Gaussianity, and only studied the effects on the power spectra arising from the lensing bispectrum. Other non-linear processes (such as the Sunyaev–Zel'dovich effect) could have larger non-Gaussianities and hence more important effects if they have not be masked or removed in the analysis.

\appendix

\section{Power spectrum from large-scale convergence and shear}
\label{app:squeezed}

For large-scale constant convergence $\kappa$ and shear $\gamma$, the magnification matrix is
$A_{ij} = (1-\kappa)\delta_{ij} - \gamma_{ij}$ with determinant $|\mA| = (1-\kappa)^2 - g^2$ where
 $g^2 \equiv \frac{\gamma_{ij}\gamma^{ij}}{2} = \gamma_{11}^2+\gamma_{12}^2$. Expanding to third order in the convergence and shear,
 and noting that $[\gamma^2]_{ij} = g^2\delta_{ij}$, we have
 \begin{multline}
 |\mA^{-1} \vell| = \ell \sqrt{\vellhat^T \mA^{-1}\mA^{-1}\vellhat} =  \biggl(
 [1 + \kappa + \lgl] + [\kappa^2 + 2\kappa\lgl + \frac{3}{2}g^2 -\frac{1}{2}(\lgl)^2] \\
 + [\kappa^3 + 3\kappa^2\lgl + \frac{1}{2} g^2(9\kappa+ \lgl) - \frac{3}{2}\kappa(\lgl)^2 + \frac{1}{2}(\lgl)^3] + \dots
 \biggr)\ell,
 \end{multline}
 where $\hat{\vell} \equiv \vell/\ell$ is a unit wavevector.
 To get the power spectrum we want to average over $\vellhat$ angles and use
 \be
 \int \ud\phi_\vell \lgl = 0, \qquad \int \frac{\ud\phi_\vell}{2\pi} (\lgl)^2 = \frac{g^2}{2}, \qquad \int \ud\phi_\vell (\lgl)^3 = 0.
 \ee
 Expanding, using the angular averages and taking the expectation value (so that $\la g^2\ra = \la\kappa^2\ra$) we then have the approximation for the lensed spectrum
\begin{multline}
\ell^2\tilde{C}_\ell^{XY}=  \int\frac{\ud\phi_\ell}{2\pi} \la \ell^2|\mA|^{-1} C^{XY}_{|\mA^{-1}\bfell|} \ra = \
\ell^2 C^{XY}_{\ell}
+\frac{1}{2}\la\kappa^2\ra \left( \frac{3}{2}\frac{\ud^2 (\ell^2C_\ell^{XY})}{\ud \ln\ell^2}  +  \frac{\ud (\ell^2C_\ell^{XY})}{d\ln\ell} \right)\\
+ \frac{1}{6}\la\kappa^3\ra \ell\frac{ \ud^3 (\ell^4C_\ell^{XY})}{\ud \ell^3}
+ \frac{1}{4}\la\kappa g^2\ra\left( \ell\frac{\ud^3 (\ell^4C_\ell^{XY})}{\ud \ell^3}
-  \frac{\ud^2 (\ell^4C_\ell^{XY})}{\ud \ell^2}  + \frac{1}{\ell} \frac{\ud (\ell^4C_\ell^{XY})}{\ud \ell}\right) +
\dots
\end{multline}
The power spectrum smoothing using this result looks very similar to Fig.~\ref{fig:squeezed} if the effective convergence variance parameter is adjusted appropriately (smaller in this case, since the shear increases the total effect). The effect of the convergence skewness is the same, but with shear there is an additional $\la g^2\kappa\ra$ term which depends on the squeezed shape of the bispectrum and hence the total effect is no longer determined by a single parameter.

\section{Gradient power spectra}
\label{Cgrads}
\begin{figure*}
\begin{center}
\includegraphics[width=180mm]{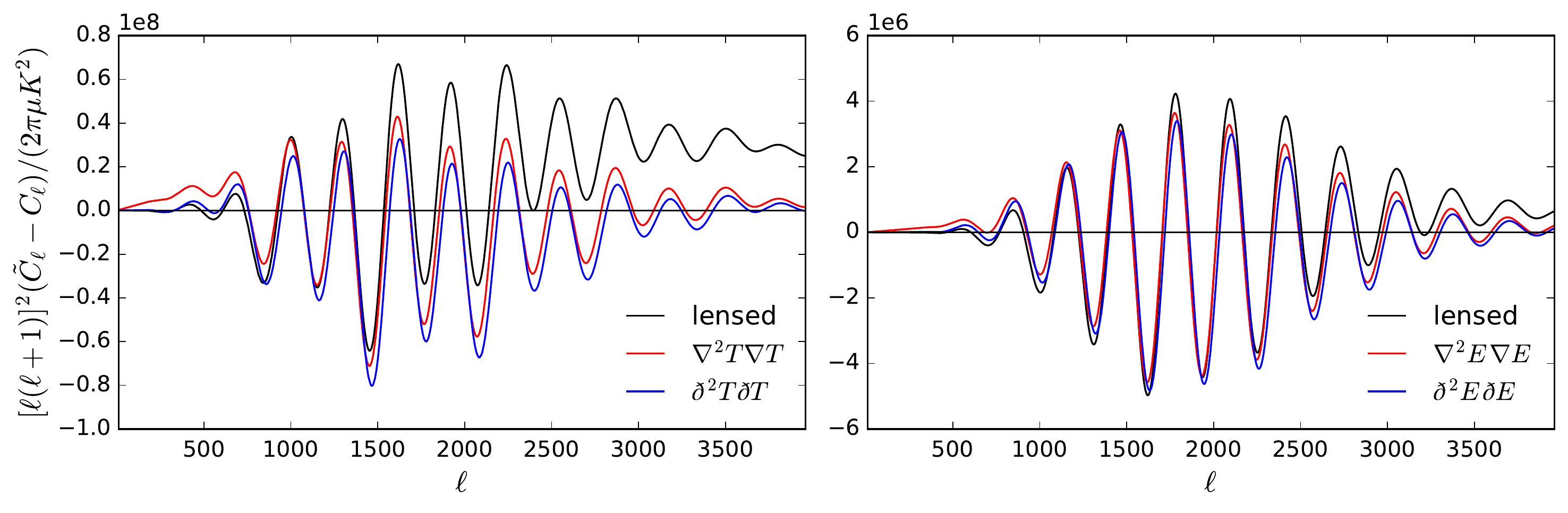}
\caption{Difference between lensed and unlensed temperature and E-polarization spectra, for the usual lensed spectrum (black) and the two gradient power spectra relevant for the response to the bispectrum (red and black; defined by Eq.~\eqref{eq:gradspecdef} and calculated in the approximation in which T and E modes are treated the same neglecting small effects at the level of the lensed B modes). The two gradient power spectra are similar,
and both are lower than the normal lensed spectrum on small scales: the gradient lensing smoothing is operating on $\ell^3$ times the normal spectrum, which reduces the transfer of power from large to small scales compared to lensing the standard spectrum. The cross-spectrum behaves similarly.
}
\label{fig:lensedspectra}
\end{center}
\end{figure*}

%
%
%
First consider the relevant lensed gradient correlation function, which can be decomposed (assuming Gaussianity of the lensing field) as
\begin{align}
\chi_{abc}^{XY}(\vr) &\equiv \la \grad_a\grad_b X(\vx+\valpha) \grad_c Y(\vx'+\valpha')\ra \nonumber\\
&= i\int \frac{\ud^2\ell}{(2\pi)^2} C_\ell^{XY} \ell_a\ell_b\ell_c e^{i\vell\cdot \vr} \exp\left(-\frac{1}{2}\ell^2[\sigma^2(r) + \cos(2\phi) \Cgtwo(r)]\right) \nonumber\\
&\equiv \chi^{XY}_3(r) \vrhat_{\la a}\vrhat_b \vrhat_{c\ra} + \frac{3}{4}\chi^{XY}_1(r) \vrhat_{(a}\delta_{bc)},
\end{align}
where $\vr = \vx-\vx'$, $\phi = \phi_{\ell}-\phi_r$, and the $\sigma^2(r)$ and $\Cgtwo(r)$ functions are as defined in the standard result for the lensing of the power spectrum~\cite{Seljak:1996ve} (our notation follows the detailed derivations in Ref.~\cite{Lewis:2006fu}). In the last line we defined correlation functions for the two irreducible tensors,
where angle brackets denote the irreducible symmetric trace free part of the enclosed indices and hats denote unit vectors.
Noting $\ell^a \ell^b \ell^c  \vrhat_{(a}\delta_{bc)} = \ell^3\cos(\phi)$ and $\vrhat_{(a}\delta_{bc)} \vrhat^{a}\delta^{bc} = \frac{4}{3}$ we have
\be
\chi^{XY}_1(r) = -\int \frac{\ud\ell}{\ell} \frac{\ell^2C_\ell^{XY}}{2\pi} e^{-\ell^2\sigma^2(r)/2}\ell^3 \left( J_1(\ell r) + \frac{\ell^2\Cgtwo(r)}{4}[J_3(\ell r) - J_1(\ell r)]+ \dots \right)
\ee
where $J_n(x)$ are Bessel functions.
Likewise using $\ell^a \ell^b \ell^c \vrhat_{\la a}\vrhat_b \vrhat_{c\ra} = \frac{\ell^3}{4}\cos(3\phi)$ we have
\be
\chi^{XY}_3(r) =
 \int \frac{\ud\ell}{\ell} \frac{\ell^2C_\ell^{XY}}{2\pi} e^{-\ell^2\sigma^2(r)/2}\ell^3 \left( J_3(\ell r) + \frac{\ell^2\Cgtwo(r)}{4}[J_1(\ell r) + J_5(\ell r)]+ \dots \right)
\ee
Transforming to harmonic space we then have
\begin{align}
\left\la \widetilde{\grad_a \grad_b X}(\vell) \widetilde{\grad_c Y}(\vell') \right\ra
  &= (2\pi)^2\delta(\vell+\vell')\int \ud \vr e^{-\vell\cdot\vr} \chi^{XY}_{abc}(\vr)\\
  &= (2\pi)^2\delta(\vell+\vell')\int \ud \vr e^{-\vell\cdot\vr} \left(\cos(3\phi)\chi^{XY}_3(r) \ell^{-3} \ell_{\la a}\ell_b\ell_{c\ra}
  +\frac{3}{4}\cos(\phi) \,\chi^{XY}_1(r) \ell^{-1} \ell_{(a}\delta_{bc)}  \right)\\
  & \equiv   i(2\pi)^2 \delta(\vell+\vell')\left(\tilde{C}^{\eth^2X\eth Y}_\ell \ell_{\la a}\ell_b\ell_{c\ra} + \frac{3\ell^2}{4} \tilde{C}^{\grad^2 X\grad Y}_\ell \ell_{(a}\delta_{bc)}  \right),
\label{eq:gradspecdef}
\end{align}
where we defined the power spectra given by
\be
\tilde{C}^{\grad^2 X\grad Y}_\ell =  - 2\pi \int r\ud r \frac{J_1(\ell r)}{\ell^3} \chi^{XY}_1(r),
\qquad
\tilde{C}^{\eth^2 X\eth Y}_\ell =   2\pi \int r\ud r \frac{J_3(\ell r)}{\ell^3} \chi^{XY}_3(r).
\ee
Numerical results for the two spectra are shown in Fig.~\eqref{fig:lensedspectra} (in the approximation in which the small-scale EE follows the scalar result derived here).

\section*{Acknowledgments}
We thank Julien Peloton for providing numerical covariances including lensing correlations.
We acknowledge support from the European Research Council under
the European Union's Seventh Framework Programme (FP/2007-2013) / ERC Grant Agreement No. [616170].



\allowdisplaybreaks


\bibliography{PostBorn,antony,cosmomc}

\end{document}